\documentclass[aoas]{imsart}

\RequirePackage{amsthm,amsmath,amsfonts,amssymb}
\RequirePackage{natbib}
\RequirePackage[colorlinks,citecolor=blue,urlcolor=blue]{hyperref}
\RequirePackage{graphicx}
\usepackage[ruled,vlined]{algorithm2e}
\usepackage{enumitem}
\usepackage{float}
\usepackage{multirow}
\usepackage{threeparttable}
\usepackage{pifont}

\usepackage{tikz}
\usetikzlibrary{shapes.geometric, intersections}
\usetikzlibrary{positioning, shapes}
\usetikzlibrary{matrix}
\usetikzlibrary{shapes.symbols}
\usepackage{pgfplots}
\usetikzlibrary{patterns,pgfplots.fillbetween}
\startlocaldefs
\numberwithin{equation}{section}
\theoremstyle{plain}
\newtheorem{lemma}{Lemma}[section]
\newtheorem{theorem}{Theorem}[section]
\newtheorem{condition}{Condition}[section]
\newtheorem{remark}{Remark}[section]

\usepackage{xcolor}

\newcommand{\vb}{\mathbf{v}}

\newcommand{\vbh}{\hat{\mathbf{v}}}
\newcommand{\xb}{\mathbf{x}}
\newcommand{\Xb}{\mathbf{X}}

\newcommand{\Pcal}{\mathcal{P}}
\newcommand{\Rbb}{\mathbb{R}}

\DeclareMathOperator*{\argmin}{arg\,min}
\DeclareMathOperator*{\argmax}{arg\,max}
\endlocaldefs

\usepackage{tcolorbox}

\begin{document}


\begin{frontmatter}
\title{Sparse group principal component analysis via double thresholding with application to multi-cellular programs}
\runtitle{SGPCA}

\begin{aug}
\author[A]{\fnms{Qi}~\snm{Xu}\ead[label=e1]{qixu@andrew.cmu.edu}},
\author[A]{\fnms{Jing}~\snm{Lei}\ead[label=e2]{jinglei@andrew.cmu.edu}}
\and
\author[A]{\fnms{Kathryn}~\snm{Roeder}\ead[label=e3]{roeder@andrew.cmu.edu}}
\address[A]{Department of Statistics and Data Science, Carnegie Mellon University
\printead[presep={\\ }]{e1,e2,e3}}

\end{aug}

\begin{abstract}
Multi-cellular programs (MCPs) are coordinated patterns of gene expression across interacting cell types that collectively drive complex biological processes such as tissue development and immune responses. While MCPs are typically estimated from high-dimensional gene expression data using methods like sparse principal component analysis or latent factor models, these approaches often suffer from high computational costs and limited statistical power. In this work, we propose Sparse Group Principal Component Analysis (SGPCA) to estimate MCPs by leveraging their inherent group and individual sparsity. We introduce an efficient double-thresholding algorithm based on power iteration. In each iteration, a group thresholding step first identifies relevant gene groups, followed by an individual thresholding step to select active cell types. This algorithm achieves a linear computational complexity of $O(np)$, making it highly efficient and scalable for large-scale genomic analyses. We establish theoretical guarantees for SGPCA, including statistical consistency and a convergence rate that surpasses competing methods. Through extensive simulations, we demonstrate that SGPCA achieves superior estimation accuracy and improved statistical power for signal detection. Furthermore, We apply SGPCA to a Lupus study, discovering differentially expressed MCPs distinguishing Lupus patients from normal subjects.
\end{abstract}


\end{frontmatter}


\section{Introduction}
Single-cell RNA sequencing (scRNA-seq) has revolutionized modern biology by enabling gene expression profiling at the resolution of individual cells. This high-resolution view reveals substantial cellular heterogeneity within complex tissues and supports annotation of cell types, where cells are grouped based on distinct transcriptional signature \citep{abdelaal2019comparison, ji2023scannotate}. Beyond identifying cell types, an emerging challenge is to understand how different cell types coordinate their gene expression within a tissue. This has motivated the study of multi-cellular programs (MCPs), which are sets of genes that are simultaneously up- or down-regulated across multiple cell types and underlie essential biological processes \citep{jerby2022dialogue, mitchel2024coordinated}. MCPs are not only biologically meaningful but also clinically relevant. For example, \citet{ling2024concerted} identified the synaptic neuron and astrocyte program (SNAP), a coordinated expression program observed across neurons and astrocytes, whose reduced activitiy is associated with schizophrenia or advanced age.

Several statistical approaches have been employed for MCP discovery, including latent factor models \citep{stegle2012using, ling2024concerted} and sparse principal component analysis (sparse PCA) \citep{witten2009penalized, jerby2022dialogue}. While PCA and factor models capture shared patterns of gene co-expression, sparse PCA is particularly suitable for genomic data by incorporating the sparsity structure. Biologically, gene expression is inherently sparse, as only a subset of genes is actively expressed in any given cell. Statistically, sparsity is crucial in the high-dimensional regime $(p>n)$, where classical PCA is known to be inconsistent without structural assumptions \citep{johnstone2009consistency}.

\begin{figure}
    \centering
    \scalebox{0.6}{
\begin{tikzpicture}[every node/.style={font=\Large}]
    \definecolor{T4color}{rgb}{0.8, 0.2, 0.2} 
    \definecolor{T8color}{rgb}{0.2, 0.8, 0.2} 
    \definecolor{NKcolor}{rgb}{0.2, 0.2, 0.8} 
    \definecolor{InactiveGene}{rgb}{0.9, 0.9, 0.9} 

    \node (T4label) at (-1.5, 0) {Cell type 1};
    \node (T8label) at (-1.5, -1) {Cell type 2};
    \node (NKlabel) at (-1.5, -2) {Cell type 3};

    \foreach \i in {1,...,10} {
        \node (Gene\i) at (2 * \i - 1, 0.8) {Gene \i};
    }

    \fill[T4color, opacity=0.9] (0.1, 0.2) rectangle (1.9, -0.2); 
    \fill[T8color, opacity=0.7] (0.1, -0.8) rectangle (1.9, -1.2); 
    \fill[NKcolor, opacity=0.5] (0.1, -1.8) rectangle (1.9, -2.2); 
    \fill[T4color, opacity=0.6] (2.1, 0.2) rectangle (3.9, -0.2); 
    \fill[InactiveGene] (2.1, -0.8) rectangle (3.9, -1.2); 
    \fill[NKcolor, opacity=0.7] (2.1, -1.8) rectangle (3.9, -2.2); 

    \fill[InactiveGene] (4.1, 0.2) rectangle (5.9, -0.2); 
    \fill[T8color, opacity=0.9] (4.1, -0.8) rectangle (5.9, -1.2); 
    \fill[InactiveGene] (4.1, -1.8) rectangle (5.9, -2.2); 

    \foreach \j in {4,...,10} {
        \fill[InactiveGene] (2 * \j-1.9, 0.2) rectangle (2 * \j-0.1, -0.2); 
        \fill[InactiveGene] (2 * \j-1.9, -0.8) rectangle (2 * \j-0.1, -1.2); 
        \fill[InactiveGene] (2 * \j-1.9, -1.8) rectangle (2 * \j-0.1, -2.2); 
    }

    \foreach \j in {1, ..., 10}{
        \draw[black] (2 * \j - 1.9, 0.2) rectangle (2 * \j - 0.1, -2.2);
    }
     
\end{tikzpicture}
}
    \caption{An illustrative example of gene co-expression pattern across three cell types. In this example, gene 1 is expressed across all cell types; gene 2 is expressed across cell types 1 and 3; gene 3 is expressed only in cell type 2. All other genes are not expressed in all cell types. Gray blocks indicate non-expression, colored blocks indicate active expression and varying transparency signifies different expression levels.}
    \label{fig: co-expression-pattern}
\end{figure}
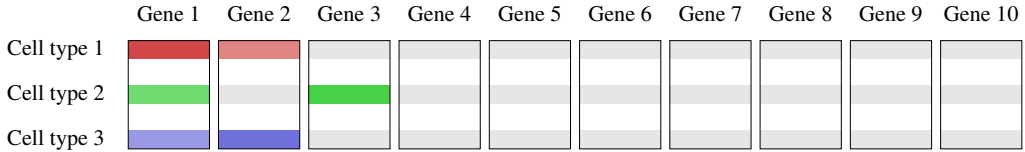

When estimating MCPs from high-dimensional, multi-cell-type gene expression data, it is important to account for a more nuanced hierarchical sparsity structure. In our setting, each gene forms a group, consisting of its expression across all cell types. Typically, only a small subset of genes participate in any given MCP, giving rise to group-level sparsity. Within each active gene, the pattern of expression may vary across cell types: a gene may be co-expressed in several, but not necessarily all, contributing cell types. This gives rise to within-group sparsity across the entries corresponding to different cell types. This hierarchical sparsity is illustrated in Figure~\ref{fig: co-expression-pattern} and is supported by biological findings in autism spectrum disorder \citep{Wamsley:2024}, aging, and schizophrenia \citep{ling2024concerted}. Accurately estimating MCPs therefore requires statistical methods that explicitly capture both levels of sparsity.

A natural approach for enforcing hierarchical sparsity arises from the sparse group lasso \citep{friedman2010note, simon2013sparse, cai2022sparse}, which uses combined $\ell_1$ and $\ell_2$ penalties to select both active groups and active variables within each group. Recently, \citet{xiao2024sparse} adapted this idea to sparse PCA using a convex relaxation based on the Fantope formulation \citep{vu2013fantope}. However, the resulting ADMM algorithm requires a costly Fantope projection step with per-iteration complexity $O(p^3)$, limiting its scalability to high-dimensional gene expression data. Moreover, the theoretical error rate $O(\sqrt{\log(p)/n})$ established in \citet{xiao2024sparse} does not exploit potential group sparsity and is therefore suboptimal in settings where hierarchical structure exists.

In this paper, we propose \textbf{Sparse Group Principal Component Analysis} (SGPCA), a scalable algorithm for estimating leading principal components (PCs) under hierarchical sparsity. Our algorithm is motivated by classical power iteration \citep{golub2013matrix} by incorporating a \emph{double-thresholding} step in each iteration: a group-wise thresholding that identifies active groups, followed by an entry-wise thresholding that promotes individual sparsity. Notably, the iterative thresholding procedure of \citep{ma2013sparse} is a special case of our approach, where only entry-wise thresholding is performed. To estimate multiple PCs, we deploy a deflation procedure \citep{jolliffe2002principal}. A key practical challenge is choosing sparsity parameters for each PC. We address this using a resampling-based criterion inspired by stability selection \citep{meinshausen2010stability, sun2013consistent}, which evaluates the alignment of estimated PCs across different resampled datasets and demonstrates strong empirical performance for variable selection.

Our method makes several contributions to the literature. First, SGPCA is computationally efficient and scalable, making it well suited for high-dimensional gene expression data. Each iteration requires $O(np)$ operations for matrix multiplication and only $O(p)$ operations for normalization and thresholding, dramatically improving upon the $O(p^3)$ computational complexity in \citep{xiao2024sparse}. Second, under a spiked covariance model with hierarchical sparsity, we establish consistency of the estimated leading PCs and derive faster convergence rates than those of \citep{xiao2024sparse} and \citep{ma2013sparse}. Third, our resampling-based tuning strategy achieves superior variable selection performance compared to commonly used criteria such as data fitting \citep{witten2009penalized} or degree of sparsity \citep{zou2006sparse}, yielding lower Type I and Type II errors in signal detection, which is particularly beneficial for MCP discovery problem. Finally, we provide an efficient and user-friendly R package implementation with an interactive tuning procedure, facilitating practical application and reproducibility. The package is released as R package \href{https://github.com/statsqixu/SGPCA}{\texttt{SGPCA}}, along with reproducible numerical experiments at \url{https://github.com/statsqixu/SGPCA_EXP}.

\subsection{Other Related Works}

Sparsity has been a central theme in high-dimensional PCA. Early methods, such as \citep{jolliffe2003modified}, imposed $\ell_1$-constraints on loading vectors but were computationally prohibitive. Subsequent approaches recast sparse PCA as a regularized regression or matrix factorization problem \citep{zou2006sparse, shen2008sparse, witten2009penalized}, enabling efficient iterative algorithms with thresholding steps. Another line of work focuses on convex relaxation of sparse PCA, and estimates PCs using semidefinite programming \citep{d2004direct, vu2013fantope}. These methods offer strong theoretical guarantees, but remain computationally demanding. For a comprehensive review, see \citep{zou2018selective}.

Tensor decomposition provides an alternative perspective by treating subjects, genes, and cell types as separate modes. Tucker decomposition has been applied to MCP discovery \citep{mitchel2024coordinated}, but consistent estimation generally requires comparable dimensionality across tensor modes \citep{zhang2018tensor}, a condition violated in typical scRNA-seq datasets where the number of cell types is much smaller than the numbers of genes or subjects. Thus, tensor-based formulations lack theoretical support in this setting.

Another related direction is multi-view learning, which integrates multiple datasets to identify shared and view-specific components. Examples include additive decompositions \citep{lock2013joint} and multiplicative structures \citep{tang2021integrated}. More recent works consider partially shared components \citep{gaynanova2019structural, lock2022bidimensional, xiao2024sparse, xu2025representation}, which flexibly model heterogeneous relationship across views. From this perspective, MCP estimation can be viewed as sparse PCA on multi-view data, where the supports of PCs across views exhibit partially sharing structure.

\subsection{Notation and Organization}

Throughout the paper, we adopt the following notation system. Random variables or vectors are denoted by capital letters such as $X$, while boldface symbols $\Xb, \xb, x$ denote a matrix, a vector, and a scalar, respectively. For any positive integer $n$, we use $[n] = \{1,2,\ldots,n\}$, and $|\cdot|$ denotes either the absolute value of a real number or the cardinality of a set. We write $a \wedge b = \min(a,b)$ and $a \vee b = \max(a,b)$. For nonnegative sequences $\{a_n\}$ and $\{b_n\}$, we write $a_n \lesssim b_n$ or $a_n = O(b_n)$ if $a_n \le C b_n$ for some constant $C>0$ independent of $n$; we write $a_n \asymp b_n$ if both $a_n \lesssim b_n$ and $b_n \lesssim a_n$ hold. Moreover, $a_n = o(b_n)$ means $a_n = c_n b_n$ for some $c_n \to 0$. For any vector $\xb$, $\|\xb\|_q$ denotes its $\ell_q$ norm. For any matrix $\Xb$, $\|\Xb\|_2 = \sup_{\|\vb\|_2 = 1}\|\Xb\vb\|_2$ denotes the spectral norm, and $\|\Xb\|_F$ denotes the Frobenius norm. By default, $\|\cdot\|$ refers to the $\ell_2$ norm for vectors and the spectral norm for matrices.

We formalize the problem setup and introduce the double thresholding algorithm and tuning parameter procedure in Section \ref{sec: method}. The consistency and convergence rate of the proposed estimator is established in Section \ref{sec: theory}. In Section \ref{sec: sim}, we present the simulation studies and evaluate estimation and signal detection performance. Furthermore, we demonstrate the discovery of multi-cellular program in Lupus data with our proposed estimator in Section \ref{sec: lupus}. Conclusions and further discussion are provided in Section \ref{sec: discussion}. 

\section{Methodology}
\label{sec: method}

Suppose our data consist of $n$ independent and identically distributed random samples, $\xb_1, \cdots, \xb_n$, from distribution from $\Pcal_X$, where each observation $\xb_i \in \Rbb^p$ follows the low-rank factor model:
\begin{align}
    \label{eqn: low-rank-model}
    \xb_i = \sum_{j=1}^{\bar{J}}\lambda_jq_{ij}\vb_j + \epsilon_i, \quad i\in[n], \quad p\gg n,
\end{align}
where vectors $\vb_1, \cdots, \vb_{\bar{J}} \in \Rbb^p$ are deterministic orthonormal loading vectors, satisfying $\vb_i\vb_j = \delta_{ij}$, $\delta_{ij} = 1$ if $i=j$ and $\delta_{ij}=0$ otherwise. The constants $\lambda_1 > \lambda_2 > \cdots > \lambda_{\bar{J}} > 0$ represent the signal strength of the corresponding loading vectors. For each sample $i$, the random vector $q_i = (q_{i1}, \cdots, q_{i\bar{J}})$ contains random scores, assumed to satisfy $\mathbb{E}[q_{ij}]=0$ and $\mathbb{E}[q_{ij}q_{i'j'}] = \delta_{ii'}\delta_{jj'}$ for all $i, i' \in [n], j, j'\in [\bar{J}]$. That is, the scores are uncorrelated across loading vectors and across samples. The noise vector $\epsilon_i \in \Rbb^p$ satisfies $\mathbb{E}[\epsilon_i] = \mathbf{0}$, $\mathbb{E}[\epsilon_i\epsilon_i^T] = \sigma^2\mathbf{I}$ and uncorrelated with $q_i$. Finally, we assume each random vector $X$ admits a known partition into $G$ disjoint groups: $[p] = \cup_{g=1}^{G}I_g$ with $I_g = \{k: k \text{ is in the $g$th group}\}$. Accordingly, $X = (X_{(I_1)}, X_{(I_2)}, \cdots, X_{(I_G)})$, $X_{(I_g)} \in \mathbb{R}^{p_g}$ and $p = \sum_{g\in[G]}p_g$. Also, each loading vector $\vb_{j}$ is partitioned conformably as $\vb_{j} = (\vb_{j, (I_1)}, \vb_{j, (I_2)}, \cdots, \vb_{j, (I_G)})$. Further, we write the entries in each group as $\vb_{j, (I_g)} = (v_{j, (I_g), 1}, \cdots, v_{j, (I_g), t}, \cdots, v_{j, (I_g), p_g})$.

Based on the above setup, the covariance matrix of $X$ is 
\begin{align}
\label{eqn: spike_cov_model}
    \Sigma = \sum_{j=1}^{\bar{J}}\lambda_j^2\vb_j\vb_j^T + \sigma^2\mathbf{I}.
\end{align}
This is the classical spiked covariance model \citep{johnstone2001distribution}: each $\vb_j$ is the $j$th eigenvector of $\Sigma$ with corresponding eigenvalue $\lambda_j^2 + \sigma^2$, and the remaining $p-\bar{J}$ eigenvalues equal $\sigma^2$. Our goal is to estimate the eigenvectors $\vb_j$, or principal components, in the this high-dimensional group-structured setting.

It is well known that in high-dimensional settings, classical PCA fails to estimate the leading eigenvectors unless additional structural assumptions are imposed \citep{johnstone2009consistency}. In this work, we adopt a \emph{hierarchical sparsity} assumption on the loading vectors $\vb_j$: for each component $j$, only a small subset of the $G$ predefined groups is active, and within each active group, $\vb_j$ is further sparse at the coordinate level. Rigorous characterization of the hierarchical sparsity will be introduced in Section \ref{sec: theory}. To exploit this two-level sparsity structure, we propose a double-thresholding algorithm, introduced in Section~\ref{subsec: dt_alg}.

In the context of MCP discovery, the low-rank factor model provides a direct and interpretable link between the statistical objects in our formulation and the underlying biological mechanisms. The random vector
$X = (X_{(I_1)}, \dots, X_{(I_G)})$ represents the expression levels of all genes across $G$ genes, where \emph{each gene corresponds to a group} and its group-level block $X_{(I_g)} \in \mathbb{R}^{T}$ contains that gene’s expression across the $T$ cell types under study. In the spiked covariance model, each loading vector $\mathbf{v}_j = (\mathbf{v}_{j,(I_1)}, \dots, \mathbf{v}_{j,(I_G)})$ defines the $j$th MCP: the block $\mathbf{v}_{j,(I_g)}$ captures how strongly gene $g$ contributes to that MCP across all cell types, and nonzero entries within $\mathbf{v}_{j,(I_g)}$ identify the specific cell types in which gene $g$ participates in the program. The gene-level group sparsity of $\mathbf{v}_j$ corresponds to the biological fact that only a subset of genes is involved in each MCP, while the within-group sparsity captures the observation that a participating gene may be active in only some of the contributing cell types. Under this interpretation, the leading eigenvectors of the covariance structure encode the dominant MCPs present in the tissue, and estimating sparse $\mathbf{v}_j$ amounts to recovering their underlying gene co-expression patterns across cell types.

\subsection{Double-Thresholding Algorithm}
\label{subsec: dt_alg}

Let $\hat{\boldsymbol{\Sigma}} = \frac{1}{n}\mathbf{X}^{\top}\mathbf{X}$ denote the sample covariance matrix. 
Without sparsity constraints, the leading principal component (PC) can be obtained via the classical power iteration algorithm \citep{golub2013matrix}, which alternates between matrix multiplication and vector normalization. 
Subsequent PCs can then be estimated by applying power iteration to the deflated covariance matrix. 

To incorporate the aforementioned hierarchical sparsity structure, we extend power iteration by introducing a \emph{double-thresholding} step. 
In each iteration, we first apply a block (group-wise) soft-thresholding operator to the groups, shrinking each group proportionally to its $\ell_{2}$ norm. 
For groups retained after this step, we then apply individual (entry-wise) soft-thresholding to induce additional sparsity across entries within each active group. The resulting algorithm is summarized in Algorithm~\ref{alg: dt}. 

\begin{algorithm}
\caption{\label{alg: dt}Double-thresholding algorithm for sparse group principal component analysis}
\KwIn{
  Sample covariance matrix $\hat{\boldsymbol{\Sigma}}$; number of principal components $J$;
  Group-wise thresholding levels $\eta_1,\ldots,\eta_J$; Entry-wise thresholding levels $\tau_1,\ldots,\tau_J$;\\
  Initial estimates $\hat{\mathbf{v}}^{(0)}_1,\ldots,\hat{\mathbf{v}}^{(0)}_J$, $\mathbf{S}=\hat{\boldsymbol{\Sigma}}$.
}
\For{$j \gets 1$ \KwTo $J$}{
  Set $k = 0$\;
  \Repeat{convergence}{
     \textbf{Multiplication}: $\hat{\boldsymbol{\gamma}}^{(k)}_j = \mathbf{S}\hat{\mathbf{v}}^{(k)}_j$\;
     \For{$g \gets 1$ \KwTo $G$}{
       \textbf{Group-wise thresholding}: 
       \[
       \hat{\boldsymbol{\gamma}}^{(k+1/2)}_{j,(g)}
       = \texttt{bst}\!\left(\hat{\boldsymbol{\gamma}}^{(k)}_{j,(g)};\, \sqrt{p_g}\,\eta_j\right)
       = \left(1 - \frac{\sqrt{p_g}\eta_j}{\|\hat{\boldsymbol{\gamma}}^{(k)}_{j,(g)}\|_2}\right)_{+}
       \hat{\boldsymbol{\gamma}}^{(k)}_{j,(g)}.
       \]
       
       \textbf{Entry-wise thresholding}: 
       \For{$t \gets 1$ \KwTo $p_g$}{
        \[
        \hat{\boldsymbol{\gamma}}^{(k+1)}_{j,(I_g),t}
        = \texttt{st}\!\left(\hat{\boldsymbol{\gamma}}^{(k+1/2)}_{j,(I_g),t};\, \tau_j\right)
        = \left(1 - \frac{\tau_j}{|\hat{\boldsymbol{\gamma}}^{(k+1/2)}_{j,(I_g),t}|}\right)_{+}
          \hat{\boldsymbol{\gamma}}^{(k+1/2)}_{j,(I_g),t}.
        \]
       }
     }
     \textbf{Normalization}: $\hat{\mathbf{v}}^{(k+1)}_j = \frac{\hat{\boldsymbol{\gamma}}^{(k+1)}_j}{\|\hat{\boldsymbol{\gamma}}^{(k+1)}_j\|_2}$\;
     $k \gets k+1$\;
  }
  \textbf{Deflation}: 
  \[
  \mathbf{S} \gets \mathbf{S} 
  - \left(\hat{\mathbf{v}}_j^{\top}\mathbf{S}\hat{\mathbf{v}}_j\right)
    \hat{\mathbf{v}}_j\hat{\mathbf{v}}_j^{\top}.
  \]
}
\KwOut{$\hat{\mathbf{v}}_1,\ldots,\hat{\mathbf{v}}_J$}
\end{algorithm}

The block soft-thresholding operator \texttt{bst} and the entry-wise soft-thresholding operator \texttt{st} are the unique minimizers of the convex penalized least squares problems
\[
\texttt{bst}(x; \lambda)
= \argmin_{y}\left\{\frac{1}{2}\|y - x\|_2^2 + \lambda\|y\|_2\right\},
\qquad
\texttt{st}(x; \lambda)
= \argmin_{y}\left\{\frac{1}{2}(y - x)^2 + \lambda|y|\right\},
\]
and alternative penalties (e.g., hard thresholding, SCAD \citealt{fan2001variable}) may also be used.

Good initialization and an appropriate stopping rule are crucial for both theoretical guarantees and empirical performance. For theoretical purpose, we construct an initialization by extending the diagonal thresholding method of \citet{johnstone2009consistency} to enforce both group- and entry-wise sparsity, as described in Algorithm~\ref{alg: modified_dt}. The algorithm stops when the estimation error achieves the desired rate comparable to the oracle estimator, as detailed in Section~\ref{sec: theory}. In practice, we initialize using SPC \citep{witten2009penalized} with coarse tuning, which consistently performs well in simulations. 
For each $\hat{\mathbf{v}}_j$, the iterations terminate at $K_j$ when the distance 
\[
d\!\left(\hat{\mathbf{v}}_j^{(K_j)}, \hat{\mathbf{v}}_j^{(K_j-1)}\right)
= \left\|
\hat{\mathbf{v}}_j^{(K_j)}\hat{\mathbf{v}}_j^{(K_j)\top}
-
\hat{\mathbf{v}}_j^{(K_j-1)}\hat{\mathbf{v}}_j^{(K_j-1)\top}
\right\|_F^2
\le \texttt{tol},
\]
which measures the Frobenius distance between the one-dimensional subspaces spanned by successive estimates and is invariant to sign flips. 
We use $\texttt{tol} = 10^{-5}$ throughout our numerical experiments.

The algorithm is computationally efficient. 
The multiplication step can be computed as $\hat{\boldsymbol{\gamma}}^{(k)}_j = \frac{1}{n}\mathbf{X}^{\top}(\mathbf{X}\hat{\mathbf{v}}^{(k)}_j)$, requiring $O(np)$ operations. Thresholding and normalization each require $O(p)$ operations, and group-wise thresholding can be parallelized across groups. Thus each iteration has complexity $O(np)$, whereas the method of \citet{xiao2024sparse} incurs $O(p^3)$ cost due to the Fantope projection. This advantage of computational efficiency makes SGPCA scalable to high-dimensional gene expression data and well suited for MCP discovery.

\subsection{Tuning parameter selection}
\label{subsec: tuning_para}

In any sparse principal component estimation method, sparsity-related tuning parameters play a crucial role in balancing the trade-off between sparsity and estimation bias. In the literature, several principles have been proposed for choosing these parameters, such as targeting a desired degree of sparsity \citep{shen2008sparse, zou2006sparse} or optimizing data-fitting accuracy \citep{witten2009penalized}. 
However, these criteria are not primarily designed for variable selection, which is of central interest in MCP discovery, where we aim to identify the genes and cell types driving each MCP.

Motivated by the stability selection principle \citep{meinshausen2010stability, sun2013consistent}, we propose a resampling-based procedure for selecting the thresholding levels in our double-thresholding algorithm, with the goal of maximizing the stability of the estimated PCs across resampled datasets. 
Specifically, given a resampling proportion $\rho \in (0,1)$ ($\rho=1/2$ in our numerical experiments), we generate $B$ subsamples by drawing $\lfloor n\rho \rfloor$ observations without replacement from the original $n$ rows of data matrix $\Xb$. For a given pair of tuning parameters $(\eta,\tau)$ and a fixed component index $j$, let $\hat{\vb}_{j,b,\eta,\tau}$ denote the estimator of the $j$th PC obtained from Algorithm~\ref{alg: dt} using the $b$th subsample $\Xb_b$. We define the alignment score for the $j$th PC under $(\eta, \tau)$ as
\[
\texttt{align}(j, \eta, \tau) 
= \frac{2}{B(B-1)} \sum_{1 \le b < b' \le B} 
\left|\hat{\vb}_{j,b,\tau,\eta}^{\top}\hat{\vb}_{j,b',\tau,\eta}\right|.
\]
Since each $\hat{\vb}_{j,b,\tau,\eta}$ is of unit norm, the absolute inner product $\left|\hat{\vb}_{j,b,\tau,\eta}^{\top}\hat{\vb}_{j,b',\tau,\eta}\right| \in [0,1]$ measures the alignment between two estimates, with the absolute value removing the sign ambiguity of eigenvectors. 
A larger value of $\texttt{align}(j,\tau,\eta)$, closer to $1$, indicates strong agreement across resamples and hence a more stable estimator for the chosen pair $(\tau,\eta)$, whereas values closer to $0$ indicate nearly orthogonal estimates and poor stability.

In Algorithm~\ref{alg: dt}, the PCs are estimated sequentially via deflation, rather than jointly as a $J$-dimensional principal subspace as in \citet{ma2013sparse}. 
This choice is driven in part by the computational burden of tuning parameter selection. 
Suppose the entry-wise thresholding levels $\tau_1,\ldots,\tau_J$ are chosen from a candidate set $\mathcal{T}$ and the group-wise thresholding levels $\eta_1,\ldots,\eta_J$ are chosen from a candidate set $\mathcal{H}$. 
Jointly estimating a $J$-dimensional principal subspace would require searching over $(|\mathcal{T}||\mathcal{H}|)^J$ combinations of tuning parameter pairs, which becomes computationally infeasible for large search spaces $\mathcal{T}$, $\mathcal{H}$ or a large number of components $J$. 
In contrast, the sequential deflation-based approach in Algorithm~\ref{alg: dt} requires only $J|\mathcal{T}||\mathcal{H}|$ combinations, leading to substantial computational savings.

On the other hand, if practitioners believe that the sparsity levels are similar across the $J$ leading PCs, it is also possible to apply the double-thresholding step within an orthogonal iteration scheme \citep{golub2013matrix} to estimate the principal subspace directly, using a common pair of thresholding levels for all components. 
In this case, Algorithm~\ref{alg: dt} can be modified to avoid the deflation step altogether. 
A further benefit of subspace-based estimation is that orthogonality among the estimated PCs is preserved, whereas deflation combined with sparsity can introduce deviations from exact orthogonality.

\begin{algorithm}[t]
\caption{\label{alg: rs}Resampling-based tuning parameter selection and PC estimation}
\KwIn{Data matrix $\Xb$; number of principal components $J$;
candidate sets $\mathcal{H}$ (group-wise thresholds) and $\mathcal{T}$ (entry-wise thresholds);
initialization $\hat{\vb}^{(0)}$; resampling proportion $\rho$; number of resamples $B$.}
\For{$j \gets 1$ \KwTo $J$}{
  \ForEach{$\eta \in \mathcal{H}$, $\tau \in \mathcal{T}$}{
    \For{$b \gets 1$ \KwTo $B$}{
      \tcp{Resampling}
      Randomly sample $\lfloor n\rho \rfloor$ rows from $\Xb$ without replacement to obtain $\Xb_b$\;
      \tcp{Estimation on resample}
      Compute $\hat{\boldsymbol{\Sigma}}_b = \frac{1}{\lfloor n\rho \rfloor}\Xb_b^{\top}\Xb_b$\;
      Apply Algorithm~\ref{alg: dt} to $\hat{\boldsymbol{\Sigma}}_b$ with tuning parameters $(\eta, \tau)$, extracting only the first PC from the deflated covariance matrix corresponding to the $j$th component. 
      Denote the resulting estimator by $\hat{\vb}_{j,b,\tau,\eta}$\;
    }
    \tcp{Alignment score}
    Compute
    \[
    \texttt{align}(j,\tau,\eta) = \frac{2}{B(B-1)}\sum_{1 \le b < b' \le B}
    \left|\hat{\vb}_{j,b,\tau,\eta}^{\top}\hat{\vb}_{j,b',\tau,\eta}\right|.
    \]
  }
  Select $(\tau_j^*, \eta_j^*) = \argmax_{\tau \in \mathcal{T}, \eta \in \mathcal{H}} \texttt{align}(j,\tau,\eta)$\;
  \tcp{Estimation on full data}
  Compute $\hat{\boldsymbol{\Sigma}} = \frac{1}{n}\Xb^{\top}\Xb$\;
  Apply Algorithm~\ref{alg: dt} to $\hat{\boldsymbol{\Sigma}}$ with tuning parameters
  $\left(\sqrt{\lfloor n\rho \rfloor / n}\,\tau_j^*,\, \sqrt{\lfloor n\rho \rfloor / n}\,\eta_j^*\right)$ 
  to estimate the $j$th PC, obtaining $\hat{\vb}_j$\;
  \tcp{Deflation}
  Update $\Xb \gets \Xb - (\Xb\hat{\vb}_j)\hat{\vb}_j^{\top}$\;
}
\KwOut{$\hat{\vb}_1,\cdots,\hat{\vb}_J$}
\end{algorithm}

The full procedure for tuning parameter selection and PC estimation is summarized in Algorithm~\ref{alg: rs}. 
From a theoretical perspective, the threshold levels $\tau_j$ and $\eta_j$ should decay at the rate $\sqrt{1/n}$ to achieve consistency, as shown in Section~\ref{sec: theory}. 
When using subsamples of size $\lfloor n\rho \rfloor$, we therefore rescale the selected tuning parameters by a factor of $\sqrt{\lfloor n\rho \rfloor / n}$ when re-estimating the PCs on the full dataset to account for the impact of sample size. 
Algorithms~\ref{alg: dt} and \ref{alg: rs} are implemented in the R package \href{https://github.com/statsqixu/SGPCA}{\texttt{SGPCA}}. 
The package also provides graphical displays of the alignment score versus the mean support size across resampled datasets, which can assist in manual inspection of the sparsity pattern and thresholding levels. 
As a byproduct, these plots are helpful for selecting the effective number of spiked components $\bar{J}$. 
Intuitively, if a PC exhibits a sparse structure, the alignment score tends to increase towards one as the mean support size approaches the true support size, and then decreases as additional noisy variables are included. 
In contrast, for $j > \bar{J}$ in the spiked covariance model \eqref{eqn: spike_cov_model}, the remaining eigenvectors are dense, and the alignment score increases monotonically with the support size. 
An illustrative example is provided in Appendix A.

\section{Statistical Property}
\label{sec: theory}

This section establishes theoretical guarantees for the estimator produced by Algorithm~\ref{alg: dt}, initialized by Algorithm~\ref{alg: modified_dt}. 
We prove consistency and provide a convergence rate under a hierarchical sparsity model that captures \emph{both} group-level and within-group sparsity. 
Compared to iterative thresholding methods designed for purely individual sparsity \citep{ma2013sparse}, our analysis accommodates grouped variables. Since Algorithm~\ref{alg: dt} estimates one leading principal component (PC) at a time, we focus on a single-spiked covariance model.

\paragraph*{Formal Setup} 
Assume $\xb_1,\ldots,\xb_n\in\mathbb{R}^p$ are generated by the single-spike model
\begin{align}
\label{eqn: single_factor_model}
    \xb_i \;=\; \lambda\, q_i\, \vb \;+\; \epsilon_i,\qquad i\in[n],
\end{align}
which is a special case of \eqref{eqn: low-rank-model} with $\bar{J}=1$ and $\sigma^2=1$.
We assume equal group size: $p_1=\cdots=p_G=T$, so that $p=GT$.
We use $n$ as the problem index, allowing $p=p(n)$, $G=G(n)$, and $T=T(n)$ to grow with $n$, while the spike strength $\lambda>0$ is treated as a constant (independent of $n$).
We impose the following distributional condition.

\begin{condition}
\label{condition_dist}
For model \eqref{eqn: single_factor_model}, we assume:
\begin{enumerate}[label=\alph*.]
    \item The latent scores $\{q_i\}_{i=1}^n$ are i.i.d.\ mean-zero, unit-variance sub-Gaussian with parameter $K_q$, i.e.
    $\mathbb{P}(|q_i|\ge t)\le 2\exp(-t^2/K_q^2)$ for all $t>0$.
    \item The noise vectors $\epsilon_i=(\epsilon_{i1},\ldots,\epsilon_{ip})$ have independent coordinates, where each $\epsilon_{ij}$ is mean-zero, unit-variance sub-Gaussian with parameter $K_\epsilon$, i.e.
    $\mathbb{P}(|\epsilon_{ij}|\ge t)\le 2\exp(-t^2/K_\epsilon^2)$ for all $t>0$.
    \item $\{q_i\}$ are independent of $\{\epsilon_{ij}\}$.
\end{enumerate}
\end{condition}

Under model \eqref{eqn: single_factor_model} and Condition~\ref{condition_dist}, the population covariance is
\[
\Sigma \;=\; \lambda^2\, \vb\vb^\top \;+\; \mathbf{I}.
\]

For unit vectors $\vb_1,\vb_2$, define the squared subspace distance
\[
d(\vb_1,\vb_2)\;:=\;\|\vb_1\vb_1^\top-\vb_2\vb_2^\top\|_F^2.
\]
Geometrically, $d(\vb_1,\vb_2)=2\sin^2\theta(\vb_1,\vb_2)$, where
$\theta(\vb_1,\vb_2)=\arccos(|\vb_1^\top \vb_2|)$ is the principal angle between the one-dimensional subspaces.
For nonzero vectors, we use the scale-invariant version
\[
d(\vb_1,\vb_2)\;:=\;\Big\|\frac{\vb_1\vb_1^\top}{\vb_1^\top\vb_1}-\frac{\vb_2\vb_2^\top}{\vb_2^\top\vb_2}\Big\|_F^2,
\]
which has the same interpretation. In addition, we consider the following high-dimensional regime:

\begin{condition}
\label{condition_gr}
As $n\to\infty$, we assume
\begin{enumerate}[label=\alph*.]
    \item $\log(G)=o(n)$ and $G\to\infty$;
    \item $T=o(\log G)$ and $T\ge 1$.
\end{enumerate}
\end{condition}

Condition~\ref{condition_gr}(a) allows the number of groups to grow up to a sub-exponential rate in $n$. Condition~\ref{condition_gr}(b) ensures $T=o(n)$. These conditions are natural in multi-cellular programming where typically $T\ll n\ll G$.

\paragraph*{Hierarchical Sparsity Assumptions}

We consider a hierarchical sparsity structure on the loading vector $\vb$.
Let $\vb_{(I_g)}\in\mathbb{R}^T$ denote the subvector of $\vb$ in group $g$. Without loss of generality, we assume groups are ordered by their magnitudes:
\[
\|\vb_{(I_1)}\|_2 \ge \cdots \ge \|\vb_{(I_G)}\|_2.
\]
We say $\vb$ belongs to the weak-$\ell_{G,r}$ ball of radius $m_G$, denoted $\vb\in w\ell_{G,r}(m_G)$, if
\begin{align}
\label{formula: group_sparsity}
    \|\vb_{(I_g)}\|_2 \le m_G\, g^{-1/r},\qquad \forall g\in[G],
\end{align}
for some fixed $0<r<2$. To formulate within-group sparsity among the leading groups, define
\[
\mathcal{V}(g)
:=\{\, v_j:\ j\in g'\ \text{for some }g'\le g \,\},
\]
i.e.\ all coordinates contained in the top $g$ groups (by $\ell_2$ norm).
Order the entries in $\mathcal{V}(g)$ by magnitude:
$|\vb_{(1)}|\ge \cdots \ge |\vb_{(gT)}|$.
We assume that for each $g$ there exists a (possibly $g$-dependent) constant $m_{(g)}$ such that
\begin{align}
\label{formula: individual_sparsity}
    |\vb_{(j)}| \le m_{(g)}\, j^{-1/r},\qquad \forall j\in[gT].
\end{align}

Throughout, we assume $r\in(0,2)$ is fixed and that $m_G$ and $\{m_{(g)}\}$ are bounded above by universal constants.
In words, group magnitudes decay quickly, and within the leading groups, individual magnitudes also decay quickly. This hierarchical sparsity is more general than simultaneous $\ell_0$-type constraints used in \citet{cai2022sparse,xiao2024sparse}, and is well aligned with applications where only a small number of groups are active and, within them, only a subset of coordinates are strong.

\paragraph*{Modified Diagonal Thresholding as Initialization} For theoretical analysis, we construct an initialization by extending the diagonal thresholding method of \citet{johnstone2009consistency} to accommodate the above hierarchical sparsity assumption, shown in Algorithm \ref{alg: modified_dt}, and specify tuning parameters as
\[
\pi_n=\pi\sqrt{\frac{\log G}{n}},
\qquad
\omega_n=\omega\sqrt{\frac{\log(T|\mathcal{B}_G|)}{n}},
\]
with sufficiently large constants $\pi,\omega>0$. 

\begin{algorithm}
\caption{\label{alg: modified_dt}Modified diagonal thresholding algorithm for initialization}
\KwIn{
  Sample covariance matrix $\hat{\Sigma}$; group threshold $\pi_n$; individual threshold $\omega_n$.
}
\KwOut{Initialization vector $\hat{\vb}^{(0)}$}
1. \textbf{Group selection}:
\[
\mathcal{B}_G = \Big\{g:\ \frac{1}{n}\sum_{i=1}^n\sum_{c\in g}x_{ic}^2 \ge T+\pi_n\Big\}.
\]
2. \textbf{Individual selection}:
\[
\mathcal{B} = \Big\{c\in \Psi(\mathcal{B}_G):\ \frac{1}{n}\sum_{i=1}^n x_{ic}^2 \ge 1+\omega_n\Big\}.
\]
3. \textbf{Reduced PCA}: compute the leading eigenvector of $\hat{\Sigma}_{\mathcal{B}\mathcal{B}}$.
\;
4. \textbf{Zero-padding}: embed this eigenvector into $\mathbb{R}^p$ by assigning zeros to all coordinates not in $\mathcal{B}$.
\end{algorithm}

\paragraph*{Key Quantities and Oracle Signal Sets}

First of all, we define
\[
h(x):=\frac{x^2}{1+x},\qquad x>0,
\]
which captures the effective signal-to-noise scaling. In particular, it behaves like $x^2$ when $x\ll 1$ and like $x$ when $x\gg 1$. Based on $h(\cdot)$, we define the group-level and entry-level \emph{loading thresholds} by
\begin{align}
\label{eq:alpha_beta_def_maintext}
\alpha_n \;=\; \alpha\Big(\frac{\log G}{n\,h(\lambda^2)}\Big)^{1/2},
\qquad
\beta_n \;=\; \beta\Big(\frac{\log(T|\mathcal{G}|)}{n\,h(\lambda^2)}\Big)^{1/2},
\end{align}
where $\alpha,\beta>0$ are fixed constants.
The corresponding oracle signal sets are
\begin{align}
\label{eq:oracle_sets_maintext}
\mathcal{G} \;=\;\{g\in[G]:\ \|\vb_{(I_g)}\|_2\ge \alpha_n\},
\qquad
\mathcal{S} \;=\;\{j\in\Psi(\mathcal{G}):\ |v_j|\ge \beta_n\}.
\end{align}
In the following lemma, we provide the upper bound of cardinality of the oracle signal sets.

\begin{lemma}
\label{lemma: cardinality_bounds}
For sufficiently large $n$, the oracle sets satisfy
\[
|\mathcal{G}|
\;\le\;
\Big(\frac{m_G}{\alpha_n}\Big)^r\wedge G,
\qquad
|\mathcal{S}|
\;\le\;
T|\mathcal{G}|
\;\wedge\;
\Big(\frac{m_{(|\mathcal{G}|)}}{\beta_n}\Big)^r.
\]
\end{lemma}

In Algorithm~\ref{alg: dt}, we choose the iteration thresholding levels to match the hierarchical signal scales:
\[
\eta_n=\eta\,\frac{\alpha_n}{\sqrt{T}},
\qquad
\tau_n=\tau\,\beta_n,
\]
with sufficiently large constants $\eta,\tau>0$.

\begin{remark}
The dependence on $|\mathcal{G}|$ in the logarithmic factor $\log(T|\mathcal{G}|)$ is used for \emph{theoretical sharpness} in the oracle analysis.
In practice, one may replace $\log(T|\mathcal{G}|)$ by $\log p$ (or tune thresholds via the resampling procedure described in Section~\ref{subsec: tuning_para}) without affecting the algorithmic implementation.
\end{remark}

\subsection{Main Results}

In the following, we present the main results of the PC estimation, where the technical details are delivered in Appendix C-E.

\begin{theorem}
\label{main_theorem}
Suppose Conditions~\ref{condition_dist} and~\ref{condition_gr} hold, and $\vb$ satisfies the hierarchical sparsity assumptions \eqref{formula: group_sparsity}--\eqref{formula: individual_sparsity}.
With sufficiently large constants $\eta,\tau,\pi,\omega$, and with $K \asymp \log n$, the estimator $\hat{\vb}^{(K)}$ produced by Algorithm~\ref{alg: dt} satisfies, for all sufficiently large $n$,
\begin{align}
\label{eq:main_rate}
d(\vb,\hat{\vb}^{(K)})
\;\lesssim\;
\Big(\Big(\frac{m_G}{\alpha_n}\Big)^r\wedge G\Big)\alpha_n^2
\;+\;
\Big(\Big(\frac{m_{(|\mathcal{G}|)}}{\beta_n}\Big)^r \wedge T\Big(\frac{m_G}{\alpha_n}\Big)^r\Big)\beta_n^2
\;+\;
\frac{\lambda^4+\lambda^2+1}{\lambda^4}\,\frac{\log G}{n},
\end{align}
with probability at least $1-C_0G^{-2}$ for a constant $C_0>0$. In particular, under the stated conditions, the right-hand side is $o(1)$ and hence $\hat{\vb}^{(K)}$ is consistent.
\end{theorem}

\paragraph*{Interpretation of the error terms.}
The bound \eqref{eq:main_rate} contains three terms with distinct interpretations.
The first term is the (group-level) nonparametric error induced by ignoring weak groups outside $\mathcal{G}$ and estimating on an effective model size of order $|\mathcal{G}|$.
The second term is the (within-group) nonparametric error induced by ignoring weak coordinates inside $\Psi(\mathcal{G})$ and estimating on an effective coordinate set of size $|\mathcal{S}|$.
The third term is a parametric term that reflects the difficulty of separating the leading spike from the bulk, and appears even without sparsity \citep{paul2012augmented}.

\begin{remark}[Reduction to the individual-sparsity case]
Consider the special case $T=1$ and $G=p$, i.e.\ each group contains a single coordinate.
Then \eqref{formula: group_sparsity} reduces to a weak-$\ell_r$ condition on the entries of $\vb$, and the within-group sparsity condition becomes redundant.
In this setting, our oracle thresholds satisfy
\[
\alpha_n=\alpha\Big(\frac{\log p}{n\,h(\lambda^2)}\Big)^{1/2},\qquad \beta_n \equiv 0,
\]
and Algorithm~\ref{alg: modified_dt} reduces to the diagonal thresholding initialization in \citet{ma2013sparse}.
Moreover, since $h(\lambda^2)$ is a constant when $\lambda$ is fixed, the resulting rate matches the classical sparse PCA rate up to constant factors in \citep{ma2013sparse}.
\end{remark}

\begin{remark}[Comparison with SDP-based approaches]
Using the inequality $\|\vb_1-\vb_2\|_2^2 \le d(\vb_1,\vb_2)\le 2\|\vb_1-\vb_2\|_2^2$, we may compare our result to rates stated in terms of $\ell_2$ error.
Under $\ell_0$-type simultaneous sparsity constraints, \citet{xiao2024sparse} obtain (up to log factors) a rate of order $\log(p)/n$ for $\|\hat{\vb}-\vb\|_2^2$.
In contrast, under hierarchical weak-$\ell_r$ structure, the first two terms in \eqref{eq:main_rate} scale as
\[
\Big(\frac{\log G}{n\,h(\lambda^2)}\Big)^{1-\frac{r}{2}}
\quad\text{and}\quad
\Big(\frac{\log(T|\mathcal{G}|)}{n\,h(\lambda^2)}\Big)^{1-\frac{r}{2}}
\]
which can be strictly smaller than $\log(p)/n$ when the hierarchical sparsity is strong (e.g.\ small $r$ and $|\mathcal{G}|\ll G$).
This reflects a statistical gain from exploiting both group-level and within-group sparsity.
\end{remark}

\section{Simulation Studies}
\label{sec: sim}

We investigate the finite-sample performance of the proposed method (SGPCA) and compare it with two widely used baselines. The first baseline is standard PCA, which ignores sparsity and is known to behave poorly in many high-dimensional regimes. The second baseline is sparse principal components (SPC; \citealt{witten2009penalized}), implemented in the \texttt{PMA} package. SPC has been broadly adopted in high-dimensional gene expression analyses; for instance, \citet{jerby2022dialogue} used SPC to identify multicellular programs from single-cell and spatial transcriptomics data.
We do not include sparse and integrative PCA (SIPCA; \citealt{xiao2024sparse}) in our numerical study because its ADMM solver is computationally intensive: in particular, the Fantope projection incurs $O(p^3)$ cost, making SIPCA impractical at the gene-expression scale. 
As reported in \citet{xiao2024sparse}, for $n=100$ and $p=350$, estimating a single PC takes roughly 30 minutes per run.

\begin{figure}[t]
    \centering
    \includegraphics[width=1.0\linewidth]{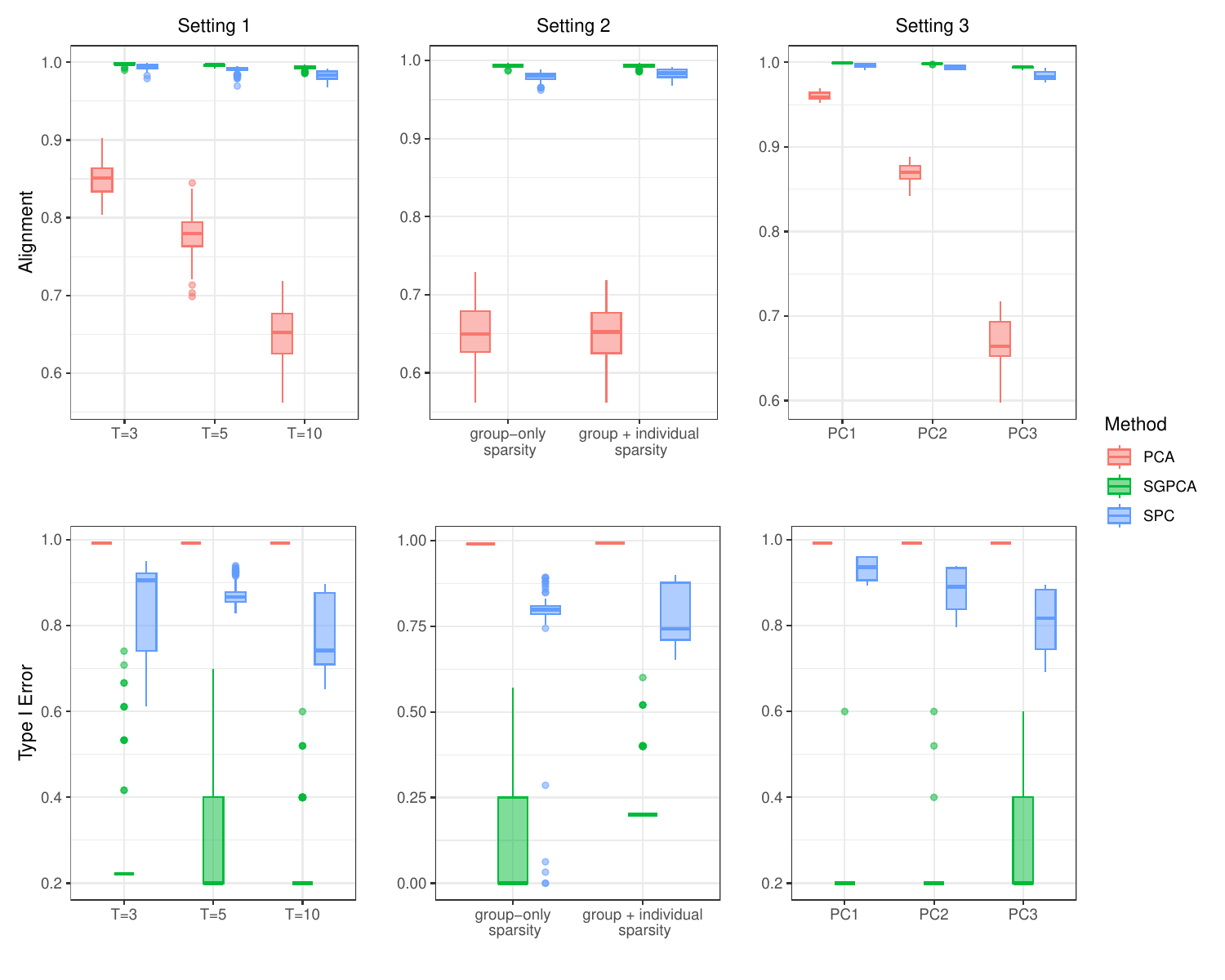}
    \caption{Simulation results for three settings. Top row: alignment. Bottom row: Type~I error.}
    \label{fig: sim-results}
\end{figure}

\paragraph*{Simulation design}
We consider three settings. In each setting, we generate $n=100$ i.i.d.\ samples from $\mathcal{N}(\mathbf{0},\Sigma)$, with $\Sigma$ specified below.

\emph{Setting 1 (varying group size; group + individual sparsity).}
We set $p = G\times T$ with $G=300$ groups and group size $T\in\{3,5,10\}$. 
The covariance is $\Sigma = 5\,\vb\vb^\top + I_p$, where $\vb$ is sparse at both the group and coordinate levels: only the first $1\%$ of groups are active, and within each active group, $80\%$ of coordinates are nonzero. 
The nonzero entries are i.i.d.\ Rademacher.

\emph{Setting 2 (different sparsity structures).}
We fix $G=300$ and $T=10$, and again use $\Sigma = 5\,\vb\vb^\top + I_p$. 
We compare two sparsity regimes: 
(i) \emph{group-only sparsity}, where all coordinates in active groups are nonzero; and 
(ii) \emph{group + individual sparsity}, where $80\%$ of coordinates in active groups are nonzero. 
Nonzero entries are i.i.d.\ Rademacher.

\emph{Setting 3 (multiple PCs).}
We simulate a multi-spiked covariance,
\[
\Sigma
=
20\,\vb_1\vb_1^\top
+
10\,\vb_2\vb_2^\top
+
5\,\vb_3\vb_3^\top
+
I_p,
\]
where each $\vb_j$ follows the same hierarchical sparsity pattern as in Setting~2(ii). 
The supports of $\vb_1,\vb_2,\vb_3$ are non-overlapping, and the nonzero entries are sampled independently from Rademacher distribution.

\paragraph*{Evaluation metrics and results.}
Figure~\ref{fig: sim-results} summarizes performance. The top row reports alignment between the estimated and true PCs (estimation accuracy), while the bottom row reports Type~I error (false positive rate; signal detection accuracy).
Across all settings, SGPCA achieves consistently high alignment, indicating accurate recovery of the underlying sparse PCs.
In Setting~1, SGPCA remains stable as group size increases, whereas SPC and PCA deteriorate substantially when $C$ is larger---a trend consistent with the known inconsistency of PCA in high-dimensional settings \citep{johnstone2009consistency}. 
In Settings~2 and~3, SGPCA attains the best estimation performance across both sparsity regimes and under multiple spikes.

For signal detection, SGPCA also yields the lowest Type~I error in every setting. 
In contrast, SPC and PCA exhibit Type~I errors close to one, indicating that they retain many noise coordinates and fail to control false positives.
Type~II errors are reported in Appendix A; SGPCA achieves Type~II error comparable to PCA and SPC, suggesting that its improved false-positive control does not come at the expense of missing true signals.

\paragraph*{Computation.}
Running times are reported in Appendix A. 
Although SGPCA is slower than PCA and SPC, it remains scalable to high-dimensional problems.
Moreover, the reported time includes an SPC-based initialization; in practice, the subsequent double-thresholding iterations are faster than the total wall-clock time suggests.
Overall, SGPCA is substantially more computationally efficient than the ADMM-based SIPCA solver \citep{xiao2024sparse}, while providing markedly better recovery of both group and individual sparsity patterns.

Finally, additional experiments with varying numbers of groups $G$ and group sizes $T$ further confirm that SGPCA consistently outperforms SPC and PCA (Appendix A).

\section{Lupus Data Example}
\label{sec: lupus}

We analyze the single-cell RNA sequencing dataset of Systemic Lupus Erythematosus (SLE), with elevated prevalence in women and individuals of Asian, African and Hispanic ancestry \citep{perez2022single}. Gene expression profiles are available in 4 major immune cell subtypes: CD4\textsuperscript{+} T (T4) cells, CD8\textsuperscript{+} T (T8) cells, Classical Monocytes (CM) cells, and Natural Killer (NK) cells, so our primary interest is discovering multi-cellular programs across these four cell subtypes related to SLE. The dataset includes 158 subjects with lupus and 98 unaffected subjects, enabling us to test whether the estimated MCPs are differentially expressed in lupus versus the normal population. 

Mathematically, discovering MCPs is equivalent to estimating PCs $\vb_1, \cdots, \vb_J$' from the covariance matrix of $\Xb$, where $J$ is the number of estimated MCPs determined by the data itself. The testing of differential expression for the $j$th MCP is formulated as follows:
\begin{align}
\label{de_testing}
    H_{0}(\vb_j): (\mu_{\text{lupus}} - \mu_{\text{normal}})^T\vb_j = 0 \quad \text{v.s.} \quad H_{1}(\vb_j): (\mu_{\text{lupus}} - \mu_{\text{normal}})^T\vb_j \neq 0,
\end{align}
where $\mu_{\text{lupus}}$ and $\mu_{\text{normal}}$ are mean gene expression levels for lupus and normal subjects, respectively. The quantity $(\mu_{\text{lupus}, i} - \mu_{\text{normal}, i})^T\vb_j$ is actually the mean difference in projection on the $j$th PC (or PC scores) between lupus and normal subjects. $H_0$ suggests that mean gene expression levels of lupus and normal subjects do not differ along the $j$th MCP, while $H_1$ indicates differential expression along the MCP between two populations. Our proposed SGPCA aims to obtain an accurate estimate of $\vb_j$' s, while the proposed debiased testing procedure \citet{zhang2024debiased} tackles the problem of double dipping. In the following, SGPCA is integrated with the debiased testing procedure to detect the differentially expressed MCPs between lupus and normal subjects.

Before estimating the MCPs and testing the differential expression, we preprocess the data using the Python package \texttt{scanpy}. For each cell type, the top 2,000 highly variable genes are selected, and further aggregated across cells from the same subject and subtype to obtain the pseudo-bulk counts for each gene, and then we remove genes expressed in less than 10 subjects. Next, we apply standard transformation, converting raw expression counts to a log normalization with a size factor (e.g., equation (2) in \citealt{ahlmann2023comparison}), which transforms a highly skewed, heavy-tailed distribution to stabilize the variance and obtain a better approximation of a normal distribution. 
In addition, we regress out library size, sex, population, and processing cohorts to remove potential confounding. Lastly, we concatenate each cell subtype matrix column-wise, as input to the Algorithm \ref{alg: dt}. The preprocessed data includes 3667 unique genes across four cell subtypes. Among these 3667 genes, 95 genes are observed in all four cell subtypes, while most genes are only recorded in one cell subtype.  

\subsection{PC Estimation}

\begin{figure}[t]
    \centering
    \includegraphics[width=1.0\linewidth]{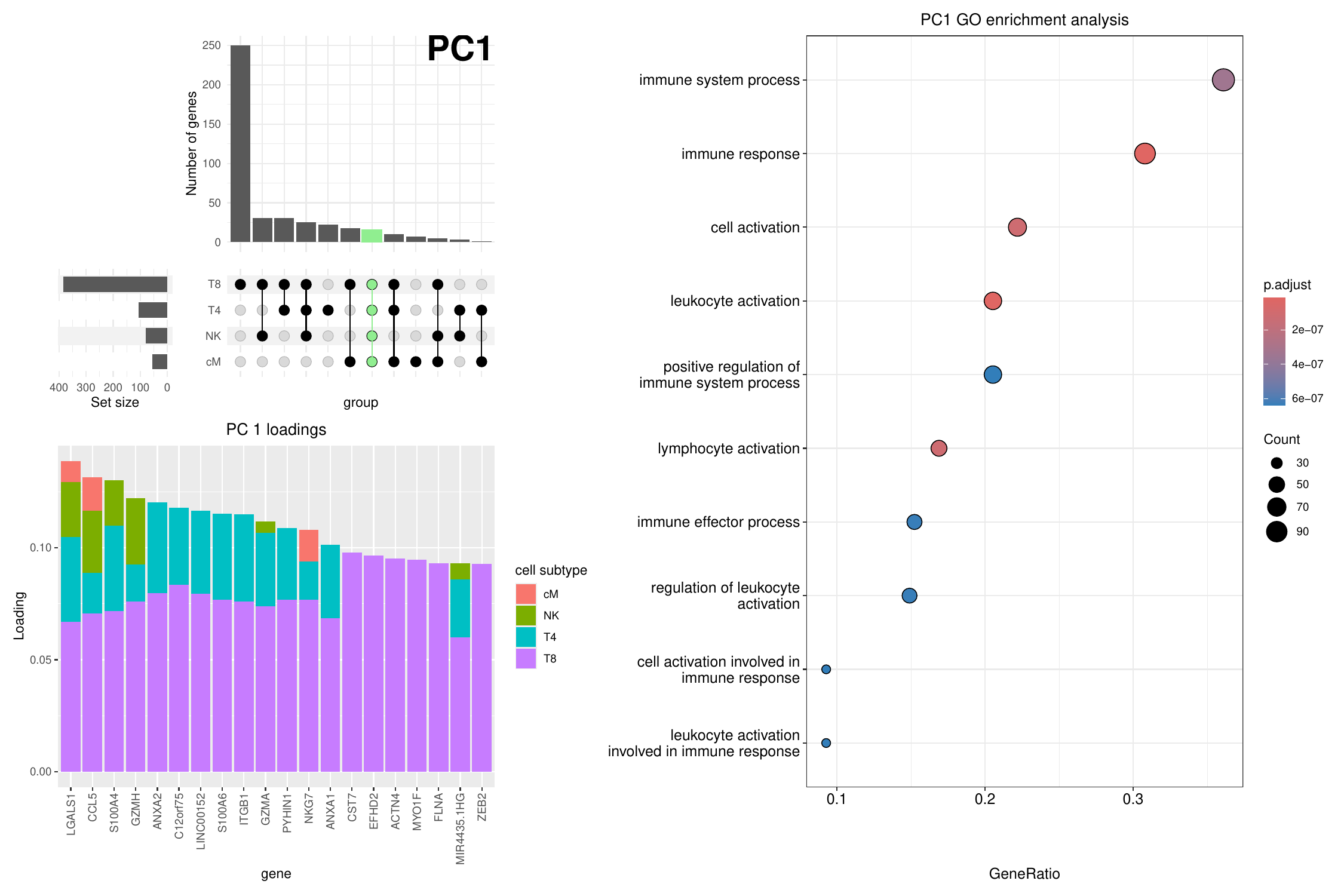}
    \caption{Estimation results for PC1. Upper left panel: a upset plot which exhibits the number of genes co-expressed across different cell subtypes. Lower left panel: Top 20 genes with the largest magnitude of loadings across subtypes in the first PC. Right panel: Top 10 biological processes enriched in the first PC.}
    \label{fig: lupus_pc1_plots}
\end{figure}

In the first step, we estimate the PCs from the processed pseudobulk gene expression matrix using SGPCA. According to graphical patterns shown in the alignment score versus the mean support size (see Figure \ref{fig: lupus_pc_tuning_plot} in Appendix B) in hyperparameter tuning, we identify $J=4$ PCs that deviate significantly from noise. In the estimated PCs, each PC detects varying number of non-zero entries, signifying co-expressed genes. Notably, many genes are co-expressed across multiple or even all cell subtypes. In particular, there are 15, 32, 18 and 45 genes co-expressed across four immune cell subtypes in the four PCs, respectively (the full list can be found in Table 6 in Appendix B).  In addition, other genes are co-expressed in two or three cell subtypes, which emphasizes the necessities of incorporating both group and individual sparsity in estimation procedure for MCPs. 

The first principal component (PC1) is primarily driven by gene expression from CD8\textsuperscript{+} T cells, suggesting their dominant contribution to inter-sample transcriptomic variability. Genes with high loadings on PC1 include key immune regulators and cytotoxic effectors such as \textit{LGALS1}, \textit{CCL5}, \textit{GZMH}, and \textit{NKG7}, indicating strong involvement of effector T cell programs. GO enrichment analysis of PC1-associated genes reveals a significant overrepresentation of terms related to leukocyte and lymphocyte activation, immune effector processes, and immune system regulation. The analysis and visualization for the other three MCPs are provided in the appendix B.

\subsection{Differential Expression Testing}

\begin{figure}
    \centering
    \includegraphics[width=1.0\linewidth]{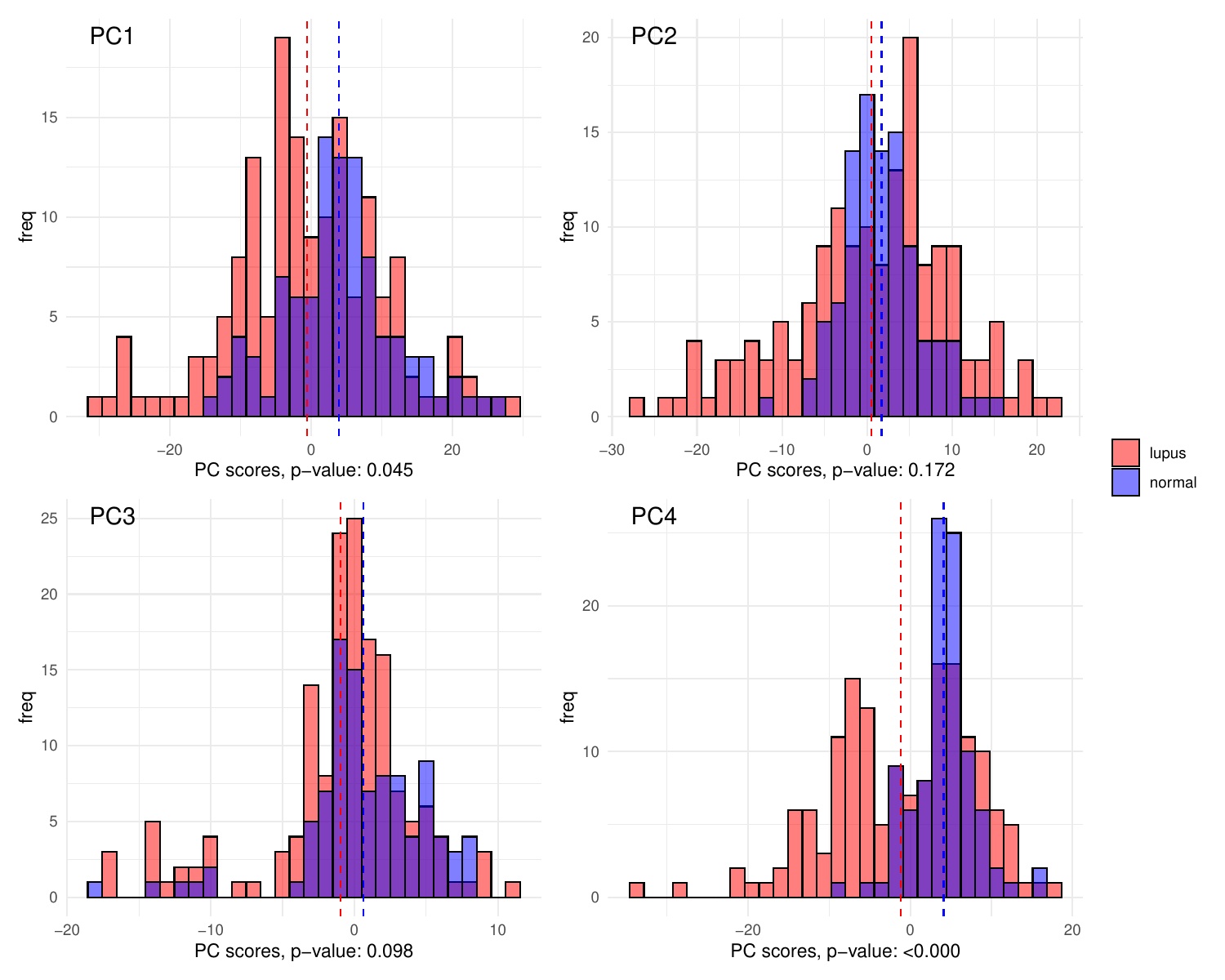}
    \caption{Histograms of scores on the 4 identified PCs. Vertical lines are score means by group. P-values are returned by debiasing test of PC scores.}
    \label{fig: lupus_test_results}
\end{figure}

Given the estimated PCs across four cell subtypes, we are interested to know whether lupus subjects have up-regulated or down-regulated expression of these MCPs compared to normal subjects. Specifically, we are performing hypothesis testing of $H_0(\vbh_j) \text{ v.s. } H_1(\vbh_j)$ for $j = 1, \cdots, 4$ where $H_0$ and $H_1$ are defined in (\ref{de_testing}). As a result, $H_0$ of PC1 and PC4 are rejected below the 0.05 threshold, indicating that gene expression profiles between lupus and normal subjects are significantly different along the direction of PC1 and PC4. In Figure \ref{fig: lupus_test_results}, the PC scores for lupus and normal subjects are visualized separately for each PC, where the PC1 and PC4 scores show visible mean differences, while the PC2 and PC3 scores are distributed similarly. Recall that PC1 is dominated by the expression of genes from CD8\textsuperscript{+} T cells, which aligns with the results of \citep{perez2022single} that CD8\textsuperscript{+} cells are functionally dysregulated in subjects with lupus.

\begin{figure}
    \centering
    \includegraphics[width=1.0\linewidth]{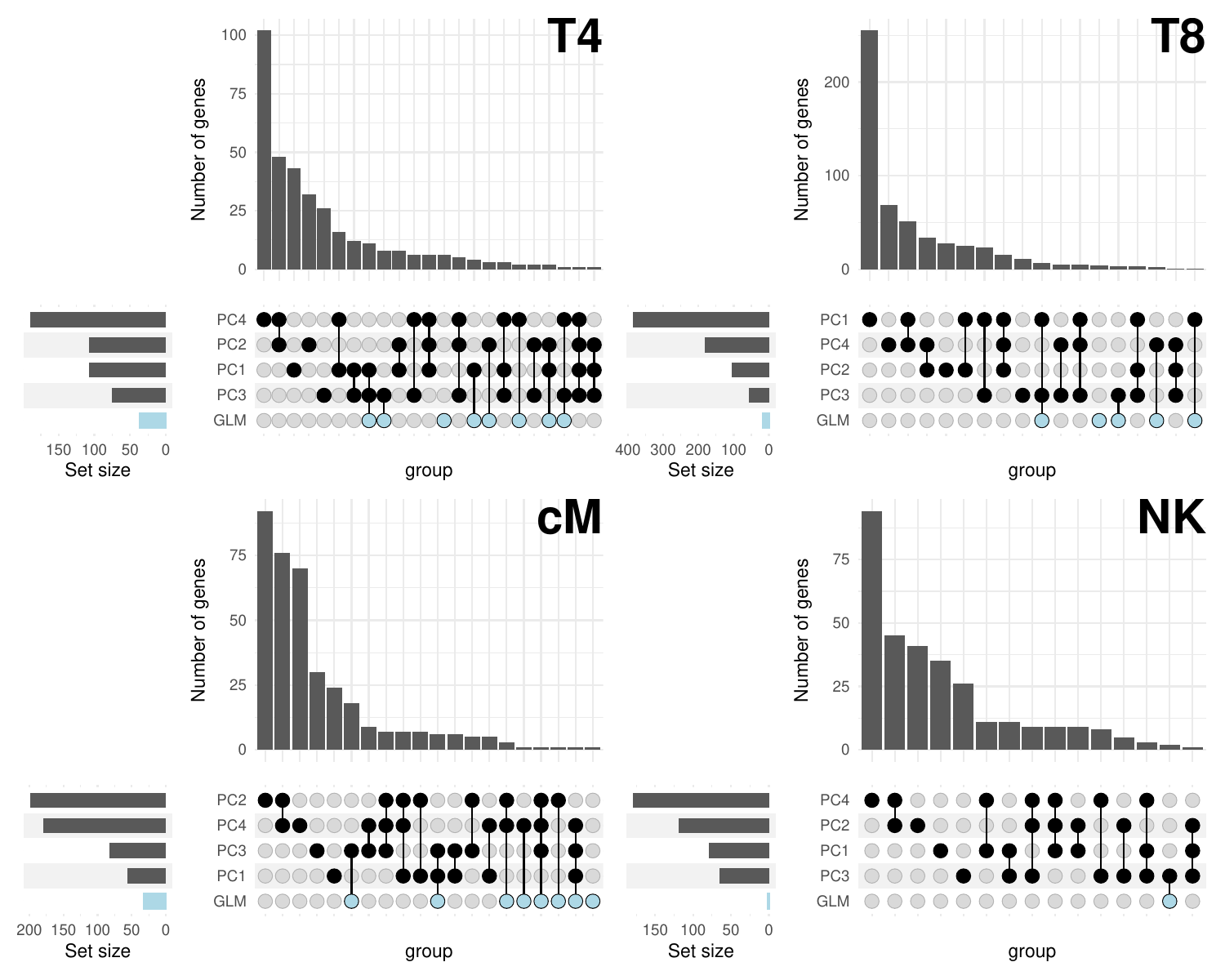}
    \caption{Upset plots of identified genes in each PC and differentially expressed genes identified by GCATE method, by cell subtypes.}
    \label{fig: gcate_compare}
\end{figure}

In addition, we compare the genes identified in each PC with the differentially expressed genes (DE) discovered using a modified GLM model  (GCATE \citep{GCATE}).   Mathematically, the debiased tests show that $(\mu_{\text{lupus}} - \mu_{\text{normal}})^T\vb \neq 0$ for PC1 and PC4, while GCATE identifies those genes with different means, that is, $I = \{i: \mu_{\text{lupus}, i} \neq \mu_{\text{normal}, i}\}$.  Thus, the objectives of the two testing strategies are fundamentally different.  The projected test aims to find a set of genes, working in concert, that differ in expression between the case and control subjects.  The marginal test aims to identify individual genes that differ in expression. For the latter test, the gene must differ substantially between cohorts to achieve significance, while many of the co-expressed genes in an MCP might have negligible differences, but these effects can add up over the program. It is interesting to note the genes in a MCP that are also marginally significant.  These may be the key drivers of the MCP that delineate the differences between the case and control subjects.  Therefore, we identify the small set $I$ in PC1 and PC4, leading to $(\mu_{\text{lupus}, I} - \mu_{\text{normal}, I})^T\vb_I \neq 0$. In Figure \ref{fig: gcate_compare}, we compare the non-zero loadings in each PC with DE genes by cell subtypes. The intersected genes between PC1, PC4 and DE genes are listed in Table \ref{tab: driver_gene_list}.

\begin{table}[H]
    \centering
    \caption{}
    \label{tab: driver_gene_list}
    \begin{tabular}{p{2cm}|p{4cm}|p{4cm}}
    \hline
    Cell subtype & PC1 & PC4 \\
    \hline
    T4 & \textit{TNFRSF4}, \textit{STAT1}, \textit{SPATS2L}, \textit{TIGIT}, \textit{TMEM244}, \textit{FOXP3}, \textit{DUSP4}, \textit{IL2RA}, \textit{IFIT3}, \textit{LAG3}, \textit{ALOX5AP}, \textit{IFI27}, \textit{ANXA2}, \textit{AC002331.1}, \textit{PMAIP1}, \textit{CD70}, \textit{IGFL2} & \textit{RTKN2}, \textit{ACVR2B}, \textit{PRDM6} \\
    \hline
    T8 & \textit{STAT1}, \textit{IFIT3}, \textit{LAG3}, \textit{OASL}, \textit{EPSTI1}, \textit{IFI27}, \textit{PATL2}, \textit{MT1E} & \textit{SLC4A10}, \textit{PRDM6} \\
    \hline
    CM & \textit{GBP1}, \textit{STAT1}, \textit{PLAC8}, \textit{IFIT2}, \textit{IFIT3}, \textit{EPSTI1}, \textit{IFI27} & \textit{IFIT2}, \textit{TNFSF10}, \textit{AREG}, \textit{GABRE}, \textit{EPPK1}, \textit{NBPF15} \\
    \hline
    NK & - & - \\
    \hline
    \end{tabular}
\end{table}

\section{Discussion}
\label{sec: discussion}

In this paper, we propose \emph{SGPCA}, a new method for estimating leading principal components under \emph{hierarchical} group- and entry-level sparsity. By explicitly leveraging both levels of sparsity, SGPCA improves estimation accuracy and signal detection, as demonstrated by our simulation studies. On the theoretical side, we establish consistency and derive a sharper convergence rate in high-dimensional asymptotic regimes.

The key intuition is simple: when the loading vector exhibits hierarchical structure, it is statistically efficient to remove noise in two stages. The first (group) screening step aggregates information within groups and filters out entire noise groups at a multiple-testing complexity proportional to $\log(G)$. Under our growth condition $T=o(\log G)$, this step incurs a mild cost while substantially reducing the ambient dimension. The second (within-group) screening step then eliminates residual noise coordinates among the surviving groups, operating at complexity $\log(T|\mathcal{G}|)$ rather than the more conservative $\log(p)$. This separation is precisely what yields the improved nonparametric component of our error bound.

From a computational perspective, each iteration of the double-thresholding refinement requires only matrix-vector multiplications and thresholding operations, and hence costs $O(np)$ time. This is substantially cheaper than approaches based on semidefinite relaxations, whose core subroutines (e.g., Fantope projections) can scale as $O(p^3)$ \citep{xiao2024sparse}. In practice, the dominant computational cost of SGPCA comes from hyperparameter selection, since each PC involves two thresholding levels. This cost can be reduced when sparsity patterns are similar across PCs, as discussed in Section~\ref{subsec: tuning_para}. Overall, SGPCA offers a statistically principled and computationally efficient approach with strong empirical performance. We also provide an efficient implementation in the R package \href{https://github.com/statsqixu/SGPCA}{\texttt{SGPCA}}.

In our application, we use SGPCA to identify multicellular programs (MCPs) in a lupus study and then perform downstream projection-based differential expression testing to detect MCPs that differ between SLE patients and healthy controls. The resulting gene sets and enriched GO terms align with biologically meaningful pathways implicated in SLE. More broadly, the simultaneous sparsity structure considered here is not specific to gene expression. For example, in online retail data, millions of products are organized into categories (e.g., electronics, clothing, groceries). Treating customers as independent samples, purchasing behavior is often sparse at both levels: most customers interact with only a small subset of categories and, within those categories, only a small subset of products. This naturally induces group- and entry-wise sparsity in latent factors. We therefore expect SGPCA to be broadly useful in other domains where a meaningful group structure is available.

\begin{acks}[Acknowledgments]
The first author would like to thank Jin-Hong Du for pre-processing the lupus dataset, and thank Tianyu Zhang for valuable discussion.
\end{acks}
\begin{funding}
This project was funded by National Institute of Mental Health (NIMH) grant R01MH123184, NSF grant DMS-2310764 and NSF grant DMS-2515687.
\end{funding}

\begin{supplement}
\stitle{Appendix A: Additional Simulation Details}
\sdescription{Additional simulation results: Type-II error and running time under simulation settings in \ref{sec: sim} and additional simulation settings.}
\end{supplement}

\begin{supplement}
\stitle{Appendix B: Real Data Analysis Details}
\sdescription{Tuning procedure for real data analysis and interpretation for extracted PC2-PC4.}
\end{supplement}

\begin{supplement}
\stitle{Appendix C-E: Technical Details for Theoretical Analysis}
\sdescription{Technical tools, proof details of theoretical results presented in Section \ref{sec: theory}.}
\end{supplement}

\bibliographystyle{imsart-nameyear} 
\bibliography{Refs}       


\end{document}